\documentclass[aps,pra,reprint,nofootinbib]{revtex4-1}
\usepackage{amsfonts,amsmath}
\usepackage{MnSymbol}
\usepackage{xspace}

\newcommand{\QKD}{QKD\xspace}
\newcommand{\Xbasis}{{\mathtt X}\xspace}
\newcommand{\Zbasis}{{\mathtt Z}\xspace}
\newcommand{\Bbasis}{{\mathtt B}\xspace}

\begin{document}
\title{Decoy State Quantum Key Distribution With More Than Three Types Of
 Photon Intensity Pulses}

\author{H. F. Chau}
\thanks{email: \texttt{hfchau@hku.hk}}
\affiliation{Department of Physics, University of Hong Kong, Pokfulam Road,
 Hong Kong}
\affiliation{Center of Theoretical and Computational Physics, University of
 Hong Kong, Pokfulam Road, Hong Kong}
\date{\today}

\begin{abstract}
 Decoy state method closes source security loophole in quantum key
 distribution (\QKD) using laser source.
 In this method, accurate estimates of the detection rates of vacuum and
 single photon events plus the error rate of single photon events are needed
 to give a good enough lower bound of the secret key rate.
 Nonetheless, the current estimation method for these detection and error
 rates, which uses three types of photon intensities, is accurate up to
 about $1\%$ relative error.
 Here I report an experimentally feasible way that greatly improves these
 estimates and hence increases the one-way key rate of the BB84 \QKD protocol
 with unbiased bases selection by at least 20\% on average in realistic
 settings.
 The major tricks are the use of more than three types of photon intensities
 plus the fact that estimating bounds of the above detection and error rates
 is numerically stable although these bounds are related to the inversion of a
 high condition number matrix.
\end{abstract}

\maketitle

 In quantum key distribution (\QKD), two trusted parties Alice and Bob share
 a secret key via preparation and measurement of photons transmitted through a
 noisy channel controlled by an eavesdropper Eve.
 Most \QKD experiments to date employ phase-randomize Poissonian distributed
 photon sources to generate photons at a reasonably high rate and use decoy
 state method to tackle Eve's photon-number-splitting attack on multiple
 photon events emitted from the Poissonian sources.
 (See, for example, Ref.~\cite{RMP09} for an overview.)
 The key idea of the decoy state method is that although Eve knows the photon
 number in each pulse, she does not know the probability distribution of
 photon number from which the pulse is drawn.
 So, by preparing each photon pulse independently from a collection of
 Poissonian sources with different intensity parameters (in other words,
 different average photon number per pulse), Alice and Bob may obtain a lower
 bound of the key rate of the final secret key they share~\cite{Wang05,LMC05}.
 Decoy state technique can handle a variety of \QKD protocols including those
 with two-way classical post-processing~\cite{MFDKTL06}, those involving the
 transmission of qudits~\cite{qudit-decoy}, and those with finite raw key
 length~\cite{LCWXZ14}.

 However, the state-of-the-art method to date, which employs a Poissonian
 source with three different types of intensities --- one high, one weak and
 one equals or closes to zero intensity, is inefficient for two reasons.
 First, using weak and zero intensity photon pulses lower the average photon
 transmission rate.
 Second, the provably secure decoy state key rate depends on good enough lower
 bounds on parameters $Y_{\Xbasis,0}$, $Y_{\Xbasis,1}$, $Y_{\Zbasis,1}$
 together with a good upper bound on $e_{\Zbasis,1}$ to be defined later.
 Nonetheless, I am going to report later in the paper that the bounds of these
 four parameters obtained through the state-of-the-art method used in
 Refs.~\cite{Wang05,LMC05,MQZL05,LCWXZ14} deviate from their actual values
 with an average relative errors of $\approx 1\%$ over a set of randomly
 chosen quantum channels.

 Here I show how to perform decoy state \QKD using higher average photon
 intensity pulses plus refined upper and lower bounds on the above four
 parameters to give a much improved provably secure key rate in realistic
 settings used in typical \QKD experiments with finite raw key length.
 The main ingredients used here are the use of more than three different types
 of photon intensities plus the observation that the lower and upper bounds
 obtained are numerically stable although one has to effectively invert a
 large condition number matrix.
 I illustrate the key idea using a specific BB84 \QKD protocol~\cite{BB84}
 which uses the $\Xbasis$ measurement results to generate the raw key and the
 $\Zbasis$ measurement results to check the channel phase error here.
 It is straightforward to extend the analysis to other \QKD schemes such as
 the six-state scheme and some qudit-based schemes~\cite{BMFBB16,Chauetal17}.

 Suppose Alice and Bob use $k\ge 2$ different photon intensities labeled
 $\mu_1 > \mu_2 > \cdots > \mu_k \ge 0$ with probabilities $p_{\mu_1}, \ldots,
 p_{\mu_k}$.
 The observed average yield per photon pulse prepared in the $\Bbasis =
 \Xbasis$ and $\Zbasis$ bases using intensity $\mu_n$ is given
 by~\cite{Wang05,LMC05,MQZL05}
\begin{equation}
 Q_{\Bbasis,\mu_n} = \sum_{m=0}^{+\infty} \frac{\mu_n^m Y_{\Bbasis,m}
 \exp(-\mu_n)}{m!} ,
 \label{E:Q_mu_def}
\end{equation}
 where $Y_{\Bbasis,m}$ is the probability of photon detection by Bob given
 that the photon pulse sent by Alice contains $m$ photons.
 (From now on, $\Bbasis$ denotes either $\Xbasis$ or $\Zbasis$.)
 Similarly, the observed average bit error rate $E_{\Bbasis,\mu_n}$ is given
 by
\begin{equation}
 Q_{\Bbasis,\mu_n} E_{\Bbasis,\mu_n} = \sum_{m=0}^{+\infty} \frac{\mu_n^m
 Y_{\Bbasis,m} e_{\Bbasis,m} \exp(-\mu_n)}{m!} ,
 \label{E:E_mu_def}
\end{equation}
 where $e_{\Bbasis,m}$ is the bit error rate for $m$~photon emission events
 prepared in the $\Bbasis$ basis.

 As $Y_{\Bbasis,m}$ and $e_{\Bbasis,m}$ are independent of the intensity
 parameter $\mu_n$ used, they can be in estimated or bounded by solving
 Eq.~\eqref{E:Q_mu_def} and~\eqref{E:E_mu_def} from a collection of intensity
 parameters $\mu_n$'s.
 Numerical stability issue aside, infinitely many intensities $\mu_n$'s are
 needed to determine all the $Y_{\Bbasis,m}$'s and $e_{\Bbasis,m}$'s.
 Fortunately, they only need to know lower bounds of $Y_{\Xbasis,0}$,
 $Y_{\Xbasis,1}$ and an upper bound of $H_2(e_{\Zbasis,1})$ to give a lower
 bound of the one-way secret key rate $R$ of the BB84 \QKD protocol with the
 raw key all coming from $\Xbasis$ measurements.
 This is because the secret key rate $R$, which is defined as the number of
 final secret bits divided by the expected number of photon pulses sent
 through the channel, is given by~\cite{LCWXZ14}
\begin{widetext}
\begin{equation}
 R = p_{\Xbasis}^2 \left\{ \langle \exp(-\mu) \rangle Y_{\Xbasis,0} + \langle
 \mu \exp(-\mu) \rangle Y_{\Xbasis,1} [1-H_2(e_p)] - \langle Q_{\Xbasis,\mu}
 H_2(E_{\Xbasis,\mu}) \rangle - \frac{\langle Q_{\Xbasis,\mu}
 \rangle}{\ell_\text{raw}} \left[ 6\log_2 \frac{\chi(k)}{\epsilon_\text{sec}}
 + \log_2 \frac{2}{\epsilon_\text{cor}} \right] \right\} .
 \label{E:key_rate}
\end{equation}
\end{widetext}
 Here $p_\Xbasis$ is the chance that Alice (Bob) uses $\Xbasis$ as the
 preparation (measurement) basis, the symbol $\langle f(\mu)\rangle$ denotes
 $\sum_{n=1}^k p_{\mu_n} f(\mu_n)$, $H_2(x) \equiv -x \log_2 x - (1-x) \log_2
 (1-x)$ is the binary entropy function, $e_p$ is the phase error rate of the
 single photon events in the raw key, and $\ell_\text{raw}$ is the length of
 the raw sifted key bits.
 For BB84, $e_p \to e_{\Zbasis,1}$ as $\ell_\text{raw} \to +\infty$.
 Also, the probability that the final secret keys shared between Alice and Bob
 are different is at most $\epsilon_\text{cor}$, Eve's information on the
 final key is at most $\epsilon_\text{sec}$~\cite{Renner05,KGR05,RGK05}, and
 $\chi(k)$ is a \QKD scheme specific factor depending on the number of photon
 intensities $k$ as well as the detailed security analysis used.
 For the case studied by Lim \emph{et al.} in Ref.~\cite{LCWXZ14}, $\chi(3) =
 21$.

 We shall see from Eq.~\eqref{E:original_Y_e_inequalities} below, the current
 method of getting an upper bound for $e_{\Zbasis,1}$ requires the knowledge
 of lower bounds of $Y_{\Zbasis,0}$ and $Y_{\Zbasis,1}$ plus an upper bound of
 $Y_{\Zbasis,1} e_{\Zbasis,1}$~\cite{MQZL05,LCWXZ14}.
 The goal, therefore, is to determine the bounds for the five parameters ---
 $Y_{\Bbasis,m}$ ($\Bbasis = \Xbasis$ and $\Zbasis$, $m=0,1$) and
 $e_{\Zbasis,1}$ --- as close to their actual values as possible using finite
 types of photon intensities $k$ (and hence finite number of
 $Q_{\Bbasis,\mu_n}$'s and $Q_{\Bbasis,\mu_n} E_{\Bbasis,\mu_n}$'s).
 The current method uses three different intensities $\mu_1 > \mu_2 > \mu_3
 \ge 0$ and the corresponding bounds are given by
\begin{subequations}
\label{E:original_Y_e_inequalities}
\begin{equation}
 Y_{\Bbasis,0} \ge \frac{\mu_2 Q_{\Bbasis,\mu_3}^{\llangle 1\rrangle}
 \exp(\mu_3) - \mu_3 Q_{\Bbasis,\mu_2}^{\llangle 0\rrangle} \exp(\mu_2)}{\mu_2
 - \mu_3} ,
 \label{E:Y0_inequality}
\end{equation}
\begin{align}
 & Y_{\Bbasis,1} e_{\Bbasis,1} \nonumber \\
 \le{} & \frac{(Q_{\Bbasis,\mu_2} E_{\Bbasis,\mu_2})^{\llangle 0\rrangle}
 \exp(\mu_2) - (Q_{\Bbasis,\mu_3} E_{\Bbasis,\mu_3})^{\llangle 1\rrangle}
 \exp(\mu_3)}{\mu_2 - \mu_3} ,
 \label{E:Y1_e1_inequality}
\end{align}
 and
\begin{align}
 Y_{\Bbasis,1} &\ge \frac{\mu_1}{\mu_1 (\mu_2-\mu_3) - \mu_2^2 + \mu_3^2}
  \left\{ Q_{\Bbasis,\mu_2}^{\llangle 1\rrangle} \exp(\mu_2) -
  Q_{\Bbasis,\mu_3}^{\llangle 0\rrangle} \times
  \vphantom{\frac{Y_{\Bbasis,0} - Q_{\mu_1}^{\llangle 0\rrangle}}{\mu_1^2}}
  \right. \nonumber \\
 & \qquad \left. \exp(\mu_3) + \frac{(\mu_2^2-\mu_3^2) [Y_{\Bbasis,0} -
  Q_{\Bbasis,\mu_1}^{\llangle 0\rrangle} \exp(\mu_1)]}{\mu_1^2} \right\}
 \label{E:Y1_inequality}
\end{align}
\end{subequations}
 provided that $\mu_1 > \mu_2 + \mu_3$~\cite{MQZL05,LCWXZ14}.
 In the above equations, $Q_{\Bbasis,\mu_n}^{\llangle i\rrangle} =
 Q_{\Bbasis,\mu_n} + (-1)^i \Delta Q_{\Bbasis,\mu_n}$ and $(Q_{\Bbasis,\mu_n}
 E_{\Bbasis,\mu_n})^{\llangle i\rrangle} = Q_{\Bbasis,\mu_n} E_{\Bbasis,\mu_n}
 + (-1)^i \Delta(Q_{\Bbasis,\mu_n} E_{\Bbasis,\mu_n})$, where $\Delta$-ed
 quantities are the upper bounds on the statistical fluctuation due to finite
 size sampling.
 From the Hoeffding inequality~\cite{Hoeffding}, these fluctuations can be
 taken to be at most
\begin{subequations}
 \label{E:Hoeffding}
\begin{equation}
 \Delta Q_{\Bbasis,\mu_n} = \frac{\langle Q_{\Bbasis,\mu} \rangle}{p_{\mu_n}}
 \left\{ \frac{\ln \left[ \frac{\chi(k)}{\epsilon_\text{sec}} \right]}{2
 s_\Bbasis} \right\}^{1/2}
 \label{E:Hoeffding_Q_mu}
\end{equation}
 and
\begin{equation}
 \Delta (Q_{\Zbasis,\mu_n} E_{\Zbasis,\mu_n}) = \frac{1}{p_{\mu_n}} \left\{
 \frac{\langle Q_{\Zbasis,\mu} \rangle \langle Q_{\Zbasis,\mu} E_{\Zbasis,\mu}
 \rangle \ln \left[ \frac{\chi(k)}{\epsilon_\text{sec}} \right]}{2 s_\Zbasis}
 \right\}^{1/2}
 \label{E:Hoeffding_Q_mu_E_mu}
\end{equation}
\end{subequations}
 each with probability at least $1-\epsilon_\text{sec}/\chi(k)$, where
 $s_\Bbasis$ is the number of bits that Alice prepares and Bob successfully
 measures in the $\Bbasis$ basis.
 Obviously, $s_\Xbasis = \ell_\text{raw}$ and $s_\Zbasis \approx
 (1-p_\Xbasis)^2 s_\Xbasis / p_\Xbasis^2$.
 Due to finite key length, the phase error rate of the single photon events in
 the raw key is upper-bounded by~\cite{FMC10}
\begin{align}
 e_p &\le e_{\Zbasis,1} + \bar{\gamma}(\epsilon_\text{sec} / \chi(k),
  e_{\Zbasis,1}, s_\Zbasis Y_{\Zbasis,1} \langle \mu \exp(-\mu) \rangle /
  \langle Q_{\Zbasis,\mu} \rangle, \nonumber \\
 & \qquad s_\Xbasis Y_{\Xbasis,1} \langle \mu \exp(-\mu) \rangle /
  \langle Q_{\Xbasis,\mu} \rangle)
 \label{E:e_p_bound}
\end{align}
 with probability at least $1-\epsilon_\text{sec}/\chi(k)$, where
\begin{equation}
 \bar{\gamma}(a,b,c,d) \equiv \sqrt{\frac{(c+d)(1-b)b}{c d} \ \ln \left[
  \frac{c+d}{2\pi c d (1-b)b a^2} \right]} .
 \label{E:gamma_def}
\end{equation}
 (This equation is based on the estimate in Ref.~\cite{LCWXZ14}, which in
 turn is deduced from Eqs.~(18) and~(22) in Ref.~\cite{FMC10}.  However, the
 factor $1/\sqrt{2\pi}$ is omitted in Eq.~(18); and the $\ln 2$ factor in
 Eq.~(22) should be in the denominator.
 The $\bar{\gamma}$ above is deducted using the same way as in
 Ref.~\cite{FMC10} with the corrected equations.
 Note further that $\bar{\gamma}$ is ill-defined if $a,c,d$ are too large.
 This is because in such case no $e_p \ge e_{\Zbasis,1}$ exists with failure
 probability $a$.
 All parameters used in this paper are carefully picked so that $\bar{\gamma}$
 is well-defined.)

 Interestingly, the $Y_{\Bbasis,0}$ bound is tight when $\mu_3\to
 0$~\cite{MQZL05,LCWXZ14}.
 (Similar bounds have been reported in Refs.~\cite{Wang05,H07}.)
 As for $\mu_1$ and $\mu_2$, they cannot be too close to each other in
 practice.
 Otherwise, the $Y_{\Bbasis,1}$ bound may not be reliable in the case of
 finite key length~\cite{Wang05}.
 In most experiments to date, the key rate $R$ is optimized by choosing $\mu_1
 \gtrsim 0.5$, $0.01 \lesssim \mu_2 \lesssim 0.1$, and $\mu_3 \approx
 0$~\cite{decoy_expt1,decoy_expt2,decoy_expt3,decoy_expt4,decoy_expt5}.

 While this choice of photon intensities can accurately determine the value
 of $Y_{\Bbasis,0}$, it is not very good at estimating $Y_{\Bbasis,1}$ and
 $e_{\Zbasis,1}$.
 Using the above photon intensities, ignoring finite key length effect, and by
 randomly picking $Y_{\Bbasis,m}$'s in $[0,1]$ and $e_{\Zbasis,m}$'s in
 $[0,0.5]$, I numerically find from Eq.~\eqref{E:original_Y_e_inequalities}
 that the average (maximum) relative errors of the estimated $Y_{\Bbasis,1}$
 and $Y_{\Xbasis,1} H_2(e_p)$ from their true values can be as high as
 $\approx 1\%$ ($\approx 5\%$).
 The deviation of the estimated $Y_{\Bbasis,1}$ from its actual value
 increases as $\mu_1$ or $\mu_2$ increase; and the deviation of the estimated
 $Y_{\Xbasis,1} H_2(e_p)$ from its actual value increases as $\mu_1$ or
 $e_{\Zbasis,1}$ increase.
 Consequently, Alice and Bob face a dilemma.
 Using small $\mu_1$ and/or $\mu_2$ give much better estimates of
 $Y_{\Bbasis,1}$ and $Y_{\Xbasis,1} H_2(e_p)$ at the expense of a lower raw
 key generation rate and hence a lower $R$.
 This is particularly true in the practical situation of a finite raw key
 length of $\lesssim 10^9$ to $10^{10}$~bits as using biased choice of photon
 intensities $p_{\mu_i}$'s cannot increase the key rate too much.

 Now, I show how to obtain a higher key rate by using a few larger $\mu_n$'s
 and a better estimates for $Y_{\Bbasis,1}$ and $Y_{\Xbasis,1} H_2(e_p)$.
 The trick is to directly solving Eqs.~\eqref{E:Q_mu_def}
 and~\eqref{E:E_mu_def} for $k\ge 2$ different photon intensities.
 I rewrite Eq.~\eqref{E:E_mu_def} as
\begin{align}
 Y_{\Bbasis,m} &= \sum_{i=1}^k (M^{-1})_{m+1,i} \left[ Q_{\Bbasis,\mu_i}
  \exp(\mu_i) - \sum_{j=k}^{+\infty} \frac{\mu_i^j Y_{\Bbasis,j}}{j!} \right]
  \nonumber \\
 &\equiv \sum_{i=1}^k (M^{-1})_{m+1,i} Q_{\Bbasis,\mu_i} \exp(\mu_i) +
  \sum_{j=k}^{+\infty} C_{m+1,j} Y_{\Bbasis,j}
 \label{E:Q_mu_variant}
\end{align}
 for $m=0,1,\ldots,k-1$, where $M_{ij} = \mu^{j-1}_i / (j-1)!$ for $1\le i,j
 \le k$.
 In this way, I may express $Y_{\Bbasis,0}, Y_{\Bbasis,1}, \ldots,
 Y_{\Bbasis,k-1}$ in terms of $Q_{\Bbasis,\mu_i}$'s for $i=1,2,\ldots ,k$,
 $Y_{\Bbasis,j}$'s for $j\ge k$, and the $k\times k$ matrix $M^{-1}$.
 Similarly, I use
\begin{align}
 Y_{\Zbasis,m} e_{\Zbasis,m} &= \sum_{i=1}^k (M^{-1})_{m+1,i}
  Q_{\Zbasis,\mu_i} E_{\Zbasis,\mu_i} \exp(\mu_i) \nonumber \\
 &\qquad + \sum_{j=k}^{+\infty} C_{m+1,j} Y_{\Zbasis,j} e_{\Zbasis,j}
 \label{E:E_mu_variant}
\end{align}
 to solve $Y_{\Zbasis,m} e_{\Zbasis,m}$ for $m=0,1,\ldots,k-1$.

 At the first glance, this straightforward approach appears to be hopeless
 because accurate bounds of $Y_{\Bbasis,m}$'s and $Y_{\Zbasis,m}
 e_{\Zbasis,m}$'s ($m<k$) require the number of photon intensities $k$ to be
 large.
 Nonetheless, large $k$ means that one has to invert the large condition
 number matrix $M$ so that the solutions of Eqs.~\eqref{E:Q_mu_variant}
 and~\eqref{E:E_mu_variant} are sensitive to perturbation on $Y_{\Bbasis,j}$'s
 and $Y_{\Zbasis,j} e_{\Zbasis,j}$'s ($j\ge k$) as well as uncertainty due to
 finite sample size for $Q_{\Bbasis,\mu_i}$'s and $Q_{\Zbasis,\mu_i}
 E_{\Zbasis,\mu_i}$'s.
 Numerical stability is an issue.
 Also, the bound on, say, $Y_{\Bbasis,1}$ requires extremization over all
 $Y_{\Bbasis,j} \in [0,1]$ for $j\ge k$, which further complicates matters.

 Upon a second thought, sensitivity to perturbation and numerical stability
 are not relevant in computing the lower bound on $R$ provided that the four
 variables $Y_{\Xbasis,0}$, $Y_{\Xbasis,1}$, $Y_{\Zbasis,1}$ and
 $Y_{\Zbasis,1} e_{\Zbasis,1}$ are insensitive to perturbation and numerically
 stable.
 I now report explicit expressions for the bounds on these four variables and
 then show that these expressions are indeed numerically stable.
% In fact, by choosing $\mu_i \le 1$, $M_{in}$ decreases exponentially with
% $n$.
% Thus, small changes in these variables lead to small changes in
% $Q_{\Bbasis,\mu_i}$ and $Q_{\Bbasis,\mu_i} E_{\Bbasis,\mu_i}$.
% As a result, the above four variables and hence the rate $R$ computed via
% Eqs.~\eqref{E:Q_mu_variant} and~\eqref{E:E_mu_variant} are not sensitive to
% perturbations provided that $\mu_1 \le 1$ (although computing all the
% $Y_{\Bbasis,j}$'s for every $j=0,1,\ldots ,k-1$ is indeed numerically
% unstable).

 Multiplying column~$i$ of the matrix $M$ by $\mu_i$ and row~$j$ by $1/(j-1)!$
 gives the corresponding Vandermonde's matrix.
 So I could use the explicit expression of the inverse of this Vandermonde's
 matrix in Ref.~\cite{Knuth} to obtain
\begin{equation}
 (M^{-1})_{m+1,i} = \frac{(-1)^{k-m-1} S_{im} m!}{\prod_{t\ne i} (\mu_i -
 \mu_t)}
 \label{E:M_inverse}
\end{equation}
 for all $0\le m\le k-1$.
 Here
\begin{equation}
 S_{im} = \sideset{}{^{'}}{\sum} \mu_{t_1} \mu_{t_2} \cdots \mu_{t_{k-m-1}} ,
 \label{E:S_im_def}
\end{equation}
 where the primed sum is over all $1\le t_1 < t_2 < \cdots < t_{k-m-1} \le k$
 with $t_1, t_2,\ldots,t_{k-m-1} \ne i$.
 Since $\{\mu_i\}_{i=1}^k$ is a strictly decreasing non-negative sequence, the
 sign of $(M^{-1})_{m+1,i}$ equals $(-1)^{k-m-i}$.
 Thus, a lower bound of $Y_{\Bbasis,m}$ is obtained by replacing
 $Q_{\Bbasis,\mu_i}$ in the R.H.S.\ of Eq.~\eqref{E:Q_mu_variant} by
 $Q_{\Bbasis,\mu_i}^{\llangle k-m-i-1 \rrangle}$; and an upper bound of
 $Y_{\Zbasis,1} e_{\Zbasis,1}$ is obtained by replacing $Q_{\Zbasis,\mu_i}
 E_{\Zbasis,\mu_i}$ in the R.H.S.\ of Eq.~\eqref{E:E_mu_variant} by
 $\left( Q_{\Zbasis,\mu_i} E_{\Zbasis,\mu_i} \right)^{\llangle k-m-i
 \rrangle}$.
 These bounds take care of the worst case deviations of measured
 $Q_{\Bbasis,\mu_i}$'s and $Q_{\Zbasis,\mu_i} E_{\Zbasis,\mu_i}$'s from their
 actual values due to finite sample size through the Hoeffding inequality.

 As for the extremization of $Y_{\Bbasis,0}$ and $Y_{\Bbasis,1}$ over
 $Y_{\Bbasis,j}$'s for $j\le k$, I need to know the signs of $C_{m+1,j}$ for
 all $j\ge k$ defined in Eq.~\eqref{E:Q_mu_variant}.
 In the Appendix, I show that $C_{1j} \ge 0$ and $C_{2j} < 0$ ($C_{1j} \le 0$
 and $C_{2j} > 0$) if $k$ is even (odd).
 In both cases, extrema occur when $Y_{\Bbasis,j} = 0$ or $1$ for all $j\ge
 k$.
 For better estimation of the four variables $Y_{\Xbasis,0}$, $Y_{\Xbasis,1}$,
 $Y_{\Zbasis,1}$ and $Y_{\Zbasis,1} e_{\Zbasis,1}$, whose values are less than
 $1/2$ in essentially all practical situations, it makes sense to bound them
 via Eqs.~\eqref{E:Q_mu_variant} and~\eqref{E:E_mu_variant} when the extrema
 occur when $Y_{\Bbasis,j} = Y_{\Zbasis,j} e_{\Zbasis,j} = 0$ for all $j\ge
 k$.
 Hence, the lower bounds of $Y_{\Bbasis,0}$ ($Y_{\Bbasis,1}$) should be found
 from Eq.~\eqref{E:Q_mu_variant} by putting $Y_{\Bbasis,j} = 0$ for all $j\ge
 k$ using $Q_{\Bbasis,\mu_i}$'s taken from even (odd) number of photon
 intensities.
 And upper bound of $Y_{\Bbasis,1} e_{\Zbasis,1}$ should be found from
 Eq.~\eqref{E:E_mu_variant} by putting $Y_{\Zbasis,j} e_{\Zbasis,j} = 0$ for
 all $j\ge k$ using $Q_{\Zbasis,\mu_i} E_{\Zbasis,\mu_i}$'s taken from even
 number of photon intensities.

\begin{table*}[th]
 \centering
 \begin{tabular}{||c|c|c|c|c|c|c||c||c|c|c|c|c|c|c||}
  \multicolumn{7}{c}{(a)} & \multicolumn{1}{c}{} & \multicolumn{7}{c}{(b)} \\
  \cline{1-7} \cline{9-15}
  & & & \multicolumn{4}{|c||}{Average key rate when $\ell_\text{raw} =$}
  & \qquad \quad {} &
  & & & \multicolumn{4}{|c||}{Average key rate when $\ell_\text{raw} =$} \\
  \cline{4-7} \cline{11-15}
  $Y_{\max}$ & & $k$ & $10^9$ & $10^{10}$ & $10^{11}$ & $\infty$
  & & $Y_{\max}$ & & $k$ & $10^9$ & $10^{10}$ & $10^{11}$ & $\infty$ \\
  \cline{1-7} \cline{9-15}
  $0.1$ & A & $3$ & $2.4\times 10^{-4}$ & $2.9\times 10^{-4}$ & $3.2\times 10^{-4}$ & $3.4\times 10^{-4}$ 
  & &
  $0.1$ & A & $3$ & $3.6\times 10^{-4}$ & $5.6\times 10^{-4}$ & $6.7\times 10^{-4}$ & $7.5\times 10^{-4}$ \\
  \cline{2-7} \cline{10-15}
  & B & $3$ & $4.4\times 10^{-4}$ & $5.1\times 10^{-4}$ & $5.4\times 10^{-4}$ & $5.6\times 10^{-4}$
  & &
  & B & $3$ & $7.6\times 10^{-4}$ & $1.0\times 10^{-3}$ & $1.2\times 10^{-3}$ & $1.3\times 10^{-3}$ \\
  \cline{2-7} \cline{10-15}
  & C & $4$ & $4.1\times 10^{-4}$ & $5.7\times 10^{-4}$ & $6.6\times 10^{-4}$ & $7.1\times 10^{-4}$
  & &
  & C & $4$ & $4.8\times 10^{-4}$ & $9.8\times 10^{-4}$ & $1.3\times 10^{-3}$ & $1.6\times 10^{-3}$ \\
  \cline{2-7} \cline{10-15}
  & D & $4$ & $5.6\times 10^{-4}$ & $6.9\times 10^{-4}$ & $7.4\times 10^{-4}$ & $7.7\times 10^{-4}$
  & &
  & D & $4$ & $8.5\times 10^{-4}$ & $1.3\times 10^{-3}$ & $1.6\times 10^{-3}$ & $1.7\times 10^{-3}$ \\
  \cline{2-7} \cline{10-15}
  & E & $5$ & $2.2\times 10^{-4}$ & $5.2\times 10^{-4}$ & $7.2\times 10^{-4}$ & $8.7\times 10^{-4}$
  & &
  & E & $5$ & $1.2\times 10^{-4}$ & $6.9\times 10^{-4}$ & $1.3\times 10^{-3}$ & $2.0\times 10^{-3}$ \\
  \cline{2-7} \cline{10-15}
  & F & $5$ & $5.3\times 10^{-4}$ & $7.3\times 10^{-4}$ & $8.2\times 10^{-4}$ & $8.9\times 10^{-4}$
  & &
  & F & $5$ & $6.7\times 10^{-4}$ & $1.3\times 10^{-3}$ & $1.7\times 10^{-3}$ & $2.0\times 10^{-3}$ \\
  \cline{2-7} \cline{10-15}
  & G & $6$ & $2.2\times 10^{-5}$ & $1.9\times 10^{-4}$ & $5.0\times 10^{-4}$ & $9.9\times 10^{-4}$
  & &
  & G & $6$ & $2.4\times 10^{-6}$ & $7.8\times 10^{-5}$ & $5.1\times 10^{-4}$ & $2.2\times 10^{-3}$ \\
  \cline{2-7} \cline{10-15}
  & H & $6$ & $2.1\times 10^{-4}$ & $5.1\times 10^{-4}$ & $7.4\times 10^{-4}$ & $9.4\times 10^{-4}$
  & &
  & H & $6$ & $1.1\times 10^{-4}$ & $5.8\times 10^{-4}$ & $1.3\times 10^{-3}$ & $2.1\times 10^{-3}$ \\
  \cline{1-7} \cline{9-15}
  $0.01$ & A & $3$ & $2.4\times 10^{-5}$ & $2.9\times 10^{-5}$ & $3.2\times 10^{-5}$ & $3.3\times 10^{-5}$
  & &
  $0.01$ & A & $3$ & $3.6\times 10^{-5}$ & $5.6\times 10^{-5}$ & $6.7\times 10^{-5}$ & $7.5\times 10^{-5}$ \\
  \cline{2-7} \cline{10-15}
  & B & $3$ & $4.5\times 10^{-5}$ & $5.1\times 10^{-5}$ & $5.4\times 10^{-5}$ & $5.6\times 10^{-5}$
  & &
  & B & $3$ & $7.6\times 10^{-5}$ & $1.0\times 10^{-4}$ & $1.2\times 10^{-4}$ & $1.3\times 10^{-4}$ \\
  \cline{2-7} \cline{10-15}
  & C & $4$ & $4.1\times 10^{-5}$ & $5.7\times 10^{-5}$ & $6.6\times 10^{-5}$ & $7.1\times 10^{-5}$
  & &
  & C & $4$ & $4.9\times 10^{-5}$ & $9.8\times 10^{-5}$ & $1.3\times 10^{-4}$ & $1.6\times 10^{-4}$ \\
  \cline{2-7} \cline{10-15}
  & D & $4$ & $5.6\times 10^{-5}$ & $6.8\times 10^{-5}$ & $7.4\times 10^{-5}$ & $7.7\times 10^{-5}$
  & &
  & D & $4$ & $8.5\times 10^{-5}$ & $1.3\times 10^{-4}$ & $1.6\times 10^{-4}$ & $1.7\times 10^{-4}$ \\
  \cline{2-7} \cline{10-15}
  & E & $5$ & $2.3\times 10^{-5}$ & $5.2\times 10^{-5}$ & $7.2\times 10^{-5}$ & $8.6\times 10^{-5}$
  & &
  & E & $5$ & $1.2\times 10^{-5}$ & $6.9\times 10^{-5}$ & $1.3\times 10^{-4}$ & $2.0\times 10^{-4}$ \\
  \cline{2-7} \cline{10-15}
  & F & $5$ & $5.3\times 10^{-5}$ & $7.3\times 10^{-5}$ & $8.2\times 10^{-5}$ & $8.8\times 10^{-5}$
  & &
  & F & $5$ & $6.7\times 10^{-5}$ & $1.3\times 10^{-4}$ & $1.7\times 10^{-4}$ & $2.0\times 10^{-4}$ \\
  \cline{2-7} \cline{10-15}
  & G & $6$ & $2.2\times 10^{-6}$ & $1.9\times 10^{-5}$ & $5.0\times 10^{-5}$ & $9.8\times 10^{-5}$
  & &
  & G & $6$ & $2.3\times 10^{-7}$ & $8.0\times 10^{-6}$ & $5.1\times 10^{-5}$ & $2.2\times 10^{-4}$ \\
  \cline{2-7} \cline{10-15}
  & H & $6$ & $2.1\times 10^{-5}$ & $5.1\times 10^{-5}$ & $7.4\times 10^{-5}$ & $9.3\times 10^{-5}$
  & &
  & H & $6$ & $1.1\times 10^{-5}$ & $5.8\times 10^{-5}$ & $1.3\times 10^{-4}$ & $2.1\times 10^{-4}$ \\
  \cline{1-7} \cline{9-15}
 \end{tabular}
 \caption{Average key rates for different decoy states with $e_{\max} = 1\%$
  for (a)~$p_\Xbasis = 0.50$ and (b)~$p_\Xbasis = 0.75$
  over a sample of $10^6$ different $Y_{\Bbasis,m}$'s and $E_{\Bbasis,m}$'s.
  For decoy parameters in case~A, $\mu_i$'s $=(0.66,0.05,10^{-6})$ and
  $p_{\mu_i}$'s $=(1/3,1/3,1/3)$.  Corresponding parameters for the other
  cases are:
  B: $(0.8,0.1,10^{-6})$ and $(1/2,1/4,1/4)$;
  C: $(0.8,0.5,0.35,10^{-6})$ and $(1/2,1/6,1/6,1/6)$;
  D: $(1.0,0.67,0.33,10^{-6})$ and $(1/2,1/6,1/6,1/6)$;
  E: $(0.8,0.65,0.5,0.35,10^{-6})$ and $(1/2,1/8,1/8,1/8,1/8)$;
  F: $(1,0.75,0.5,0.1,10^{-6})$ and $(1/2,1/8,1/8,1/8,1/8)$;
  G: $(1,0.8,0.65,0.5,0.35,10^{-6})$ and $(0.5,0.1,0.1,0.1,0.1,0.1)$; and
  H: $(1,0.8,0.6,0.4,0.2,10^{-6})$ and $(0.5,0.1,0.1,0.1,0.1,0.1)$.
  \label{T:1}}
\end{table*}

 To obtain a better bound, more photon intensities can be used.
 To obtain a higher key rate $R$, some of the photon intensities should be as
 high as $1$.
 (If $\mu > 1$, the chance of having multiple photon event is too high that
 $R$ is compromised.)
 Nevertheless, deviations between the actual and measured values of
 $Q_{\Bbasis,\mu_i}$'s increase as more photon intensities $k$ is used to
 obtain a raw key of a given length $\ell_\text{raw}$.
 These deviations may further amplify by $M^{-1}$ in
 Eqs.~\eqref{E:Q_mu_variant} and~\eqref{E:E_mu_variant}.
 My numerical finding suggests that four to five photon intensities seem to
 give optimal key rates using realistic parameters.
 Out of the $k$ photon intensities, I use data from the least $2\lfloor
 k/2\rfloor$ of them to obtain the lower bound of $Y_{\Xbasis,0}$ and the
 upper bound of $Y_{\Zbasis,1} e_{\Zbasis,1}$.
 And I use data from the least $2\lfloor (k-1)/2\rfloor+1$ photon intensities
 to obtain the lower bound of $Y_{\Bbasis,1}$.
 To sum up, the bounds I use are
\begin{subequations}
\label{E:final_bounds}
\begin{equation}
 Y_{\Bbasis,0} \ge \max \left( 0, \sum_{i=k_0}^k
 \frac{-Q_{\Bbasis,\mu_i}^{\llangle k_0-i \rrangle} \exp[\mu_i]
 \hat{\prod}_{j\ne i} \mu_j}{\hat{\prod}_{t\ne i} [\mu_i - \mu_t]} \right) ,
 \label{E:final_Y0_bound}
\end{equation}
\begin{equation}
 Y_{\Bbasis,1} \ge \max \left( 0, \sum_{i=3-k_0}^k
 \frac{-Q_{\Bbasis,\mu_i}^{\llangle k_0-i \rrangle} \exp[\mu_i]
 \hat{S}_i}{\hat{\prod}_{t\ne i} [\mu_i - \mu_t]} \right)
 \label{E:final_Y1_bound}
\end{equation}
 and
\begin{equation}
 Y_{\Zbasis,1} e_{\Zbasis,1} \le \min \left( \frac{1}{2}, \sum_{i=k_0}^k
 \frac{\left[ Q_{\Zbasis,\mu_i} E_{\Zbasis,\mu_i}
 \right]^{\llangle k_0-i \rrangle} \exp[\mu_i] \hat{S}_i}{\hat{\prod}_{t\ne i}
 [\mu_i - \mu_t]} \right) ,
 \label{E:final_e1_bound}
\end{equation}
\end{subequations}
 where $k_0 = 1 (2)$ if $k$ is even (odd), and $\hat{\prod}_{t\ne i}$ is over
 the dummy variable $t$ from $k_0$ to $k$ but skipping $i$.
 Besides, $\hat{S}_i = \sum'' \mu_{t_1} \mu_{t_2} \cdots \mu_{t_{k-k_0-1}}$
 where the double primed sum is over $k_0 \le t_1 < t_2 < \cdots < t_{k-k_0-1}
 \le k$ with $t_1,t_2,\dots,t_{k-k_0-1} \ne i$.
 (I have added the trivial conditions in Eq.~\eqref{E:final_bounds} to prevent
 the variables used in Eq.~\eqref{E:key_rate} from taking on absurd values.)

 Interestingly, this method reduces to the bounds in
 Eq.~\eqref{E:original_Y_e_inequalities} in the case of $k=3$ and $\mu_3 = 0$.
 More importantly, it is obvious from Eq.~\eqref{E:final_bounds} that these
 bounds are numerically stable provided that the photon intensities $\mu_i$'s
 are not close, say, with differences of at least $1/10$ so that the lost of
 precision in $1/(\mu_i - \mu_t)$ is not serious even taken the intensity
 fluctuation in realistic source into consideration.
 (Intensity fluctuation of order of $10^{-2}$ is easily attained in real
 experiments using a strong intensity and power stable laser plus suitable
 attenuators.)
 In contrast, it is clear that the $m!$ factor in Eq.~\eqref{E:M_inverse} is
 the origin of the numerical instability of finding $Y_{\Bbasis,m}$'s for
 large $m$'s.

 I follow Ref.~\cite{LCWXZ14} by using the following security parameters:
 $\epsilon_\text{cor} = 10^{-15}$ and $\epsilon_\text{sec} = \kappa
 \ell_\text{final}$, where $\ell_\text{final} \approx R s_\Xbasis /
 (p_\Xbasis^2 \langle Q_{\Xbasis,\mu} \rangle)$ is the length of the final key
 and $\kappa = 10^{-15}$ can be interpreted as the secrecy leakage per final
 secret bit.
 Following the derivation in Ref.~\cite{LCWXZ14}, $\chi(k) = 9 + (4k-2)$.
 (The term $4k-2$ comes from $2\times 2\lfloor k/2\rfloor + 2\times [2\lfloor
 (k-1)/2\rfloor+1]$, and the term $9$ is independent of the number of photon
 intensities used.  Note that using this method, $\chi(3) = 19$ which is less
 than the $\chi$ used by the current method~\cite{LCWXZ14} by $2$.
 It gives a slightly higher $R$.)

 I first study the key rate on a dedicated 100~km long optical fiber system
 using the channel model in Ref.~\cite{LCWXZ14}, whose channel parameters are
 deduced from the experiment in Ref.~\cite{WLGHZG12}.
 In this system, $Q_{\Bbasis,\mu} \approx (1+p_\text{ap})(2p_\text{dc}+
 \eta_\text{sys}\mu)$ and $Q_{\Bbasis,\mu} E_{\Bbasis,\mu} \approx (1+
 p_\text{ap}) p_\text{dc} + (e_\text{mis} \eta_\text{ch} + p_\text{ap}
 \eta_\text{sys} / 2)\mu$ for $0\le \mu\le 1$, with after pulse probability
 $p_\text{ap} = 4\times 10^{-2}$, dark count probability $p_\text{dc} =
 6\times 10^{-7}$, error rate of the optical system $e_\text{mis} = 5\times
 10^{-3}$, transmittances of the fiber and the system $\eta_\text{ch} = 1
 \times 10^{-2}$ and $\eta_\text{sys} = 1\times 10^{-3}$~\cite{LCWXZ14}.
 Fixing $s_\Xbasis = 10^9$ and the minimum photon intensity to $1\times
 10^{-6}$, while optimizing over $p_\Xbasis$ as well as all other photon
 intensities $\mu$'s and all the $p_\mu$'s, I find that the optimized one-way
 key rates for using $k=3,4,5$ equal $1.51\times 10^{-5}$, $1.57\times
 10^{-5}$ and $1.46\times 10^{-5}$, respectively.
 That is to say, using $k=4$ in this case increases the key rate by more than
 about $4\%$ over the standard $k=3$ method.

 To further study the general performance on different channels, I compute the
 average key rate $\langle R\rangle$ over a random sample of uniformly and
 independently distributed $Y_{\Bbasis,m} \in [0,Y_{\max}]$ and $e_{\Bbasis,m}
 \in [0,e_{\max}]$ for all $m\ge 0$.
 (But I set $e_{\Bbasis,0} = 1/2$ as this is the basic assumption on the
 detector used.  I also set $R=0$ for those sample channels whose key rates
 from Eq.~\eqref{E:key_rate} are negative.)
 Table~\ref{T:1} shows $\langle R\rangle$'s for various choices of $\mu_i$'s,
 $p_{\mu_i}$'s, $Y_{\max}$ and $\ell_{\max}$ using either unbiased or biased
 bases selection when $e_{\max} = 1\%$.
 (Note that the intensities and probabilities of the two $k=3$ cases in the
 table are adapted from actual experiments~\cite{decoy_expt4,decoy_expt5}.)
 As expected, the general trend is that the higher the value of $k$, the
 higher the average key rate in the infinite $\ell_{\max}$ limit.
 The increase in $\langle R\rangle$ by using six photon intensities can be as
 high as $77\%$.
 Whereas for $\ell_\text{raw} = 10^9$ ($10^{10}$), using four (five) types of
 photon intensities performs better because finite-size fluctuations on
 $Q_{\Bbasis,\mu_i}$'s and $E_{\Bbasis,\mu_i}$'s are relatively smaller.
 The increase in $\langle R\rangle$ in these cases is at least $26\%$ ($12\%$)
 using unbiased (biased) bases selection.
 Among the cases with the same $k$, Table~\ref{T:1} suggests that those
 using evenly distributed $\mu_i$'s in $[0,1]$ in general have a slightly
 higher $\langle R\rangle$.
 It is instructive to know why.
 Lastly, I find that the relative errors of the bounds $Y_{\Bbasis,1}$ and
 $Y_{\Zbasis,1} e_{\Zbasis,1}$ from their actual values for cases~A--H reduces
 from $\approx 10^{-2}$ to $\approx 10^{-4}$ when $k$ increases from $3$ to
 $6$ in the infinite raw key length limit.
 An explanation is given in the Appendix.

 In summary, I demonstrated the effectiveness of using more than three photon
 intensities, with several close to $1$ intensities used with significant
 chance, to obtain a high provably secure key rate through tighter bounds on
 $Y_{\Bbasis,1}$ and $e_{\Zbasis,1}$ and at the same time a higher value of
 $\langle \mu \exp (-\mu)\rangle$.
 Initial study here shows an average of at least $20\%$ improvement on the
 average key rate.
 It is instructive to further optimize the choice of intensity parameters
 $\mu_i$'s and $p_{\mu_i}$'s to see how far one can go.

\appendix
\section{The Signs Of $\boldsymbol{C_{0j}}$ And $\boldsymbol{C_{1j}}$}
\label{App:C_ji_properties}
 From Eq.~\eqref{E:M_inverse},
\begin{equation}
 C_{m+1,j} = \frac{(-1)^{k-m} m!}{j!} \sum_{i=1}^k \frac{\mu_i^j
 S_{im}}{\prod_{t\ne i} (\mu_i - \mu_t)}
 \label{E:M_inverse_product}
\end{equation}
 for all $0\le m\le k-1$.
 From Eq.~\eqref{E:S_im_def}, only the first two terms in the above sum
 contain the factor $(\mu_1 - \mu_2)$ in their denominators.
 And by summing these two terms, this factor is canceled if $j\ge 0$.
 Note further that $C_{m+1,j}$ is a symmetric function of variables $\mu_i$'s.
 Hence, $C_{m+1,j}$ is a homogeneous polynomial of degree~$\le j-m$ in the
 case of $j\ge m$.
 If $j\ge 1$, terms in the degree~$j$ homogeneous polynomial $C_{1j}$ contain
 the common factor $\prod_{n=1}^k \mu_n$.
 Therefore, $C_{1j}$ is actually a constant whenever $1\le j < k$.
 Whereas for $j\ge k$, by counting the leading power term for $\mu_1$ in
 $C_{1j}$, I conclude that $C_{1j}$ is a homogeneous polynomial of degree~$j$.
 Then, by fixing $\mu_2, \mu_3, \ldots, \mu_k$ and considering the series
 expansion of $1/(\mu_1 - \mu_n)$'s in the large $\mu_1$ limit, I get
\begin{align}
 C_{1j} &= \frac{(-1)^k}{j!} \left( \prod_{t=1}^k \mu_t \right) \left[
  \mu_1^{j-k} \prod_{r=2}^k \left( 1 + \frac{\mu_r}{\mu_1} +
  \frac{\mu_r^2}{\mu_1^2} + \cdots \right) \right. \nonumber \\
 &\qquad \left. \vphantom{\prod_{r=2}^k \frac{\mu_r^2}{\mu_1^2}} +
  f(\mu_2,\mu_3,\ldots,\mu_k) \right]
 \label{E:C1_expansion}
\end{align}
 for some function $f$ independent of $\mu_1$.
 As $C_{1j}$ is a homogeneous polynomial, by equating terms in powers of
 $\mu_1$, I arrive at
\begin{equation}
 C_{1j} = \frac{(-1)^k}{j!} \left( \prod_{t=1}^k \mu_t \right)
 \sum_{\substack{t_1+\cdots + t_k = j-k, \\
  t_1,\ldots,t_k\ge 0}} \mu_1^{t_1} \mu_2^{t_2} \cdots \mu_k^{t_k}
 \label{E:C1_result}
\end{equation}
 for all $j\ge k$.
 As all $\mu_i$'s are non-negative, I conclude that $C_{1j} \ge 0$ if $k$ is
 even and $C_{1j} \le 0$ if $k$ is odd with equality holds if and only if the
 least photon intensity $\mu_n = 0$.

 The same argument leads to
\begin{align}
 C_{2j} &= \frac{(-1)^{k-1}}{j!} \left( \sum_{t=1}^k \mu_1 \cdots \mu_{t-1}
  \mu_{t+1} \cdots \mu_k \right) \times \nonumber \\
 &\qquad \sum_{\substack{t_1+\cdots + t_k = j-k, \\
  t_1 > 0, t_2,\ldots,t_k\ge 0}} \mu_1^{t_1} \mu_2^{t_2} \cdots \mu_k^{t_k} +
  f'(\mu_2,\mu_3,\ldots,\mu_k)
 \label{E:C2_result}
\end{align}
 for all $j\ge k$.
 Here the function $f'$ can be found by recursively expanding
 Eq.~\eqref{E:M_inverse_product} in the same way as Eq.~\eqref{E:C1_expansion}
 in powers of $\mu_2$ but with $\mu_1$ set to $0$, and then in powers of
 $\mu_3$ with $\mu_1$ and $\mu_2$ set to $0$, and so on.
 Although the resultant expression is very complicated, it is easy to see that
 $C_{2j} < 0$ if $k$ is even and $C_{2j} > 0$ if $k$ is odd.
 Interested readers may apply this method to find explicit expressions for
 $C_{m+1,j}$'s for $m>2$.

 Finally, suppose the bound of $Y_{\Bbasis,1}$ or $Y_{\Zbasis,1}
 e_{\Zbasis,1}$ is obtained from a set $K$ of $k'$ different photon
 intensities.
 Then, the relative error of the bound of $Y_{\Bbasis,1}$ from its actual
 value is about $|C_{2k'}|$.
 If one of the used intensities $\mu_r \approx 0$, then
 Eq.~\eqref{E:C2_result} gives $|C_{2k'}| \approx \prod_{t\in K\setminus
 \{r\}} \mu_t / k'!$.
 For the choice of parameters in Table~\ref{T:1}, $|C_{2k'}|$ reduces from
 about $2\%$ to $0.03\%$, which is consistent with the numerical findings.
 More importantly, this estimation justifies the use of the least $2\lfloor
 k/2\rfloor$ or $2\lfloor (k-1)/2\rfloor + 1$ intensities to bound the
 variables in the main text.

\begin{acknowledgments}
 This work is supported by the RGC grant~17304716 of the Hong Kong SAR
 Government.
 I would like to thank X.D.\ Cui and Z.-Q.\ Yin for their discussions on laser
 intensity fluctuation.
\end{acknowledgments}

\bibliographystyle{apsrev4-1}

\bibliography{qc74.4}

\end{document}